\documentclass[10pt,aps,prl,twocolumn,floatfix,superscriptaddress,showpacs]{revtex4-1}

\usepackage{graphicx} 
\usepackage{amsmath} 
\usepackage{amssymb} 

\usepackage{color}

\begin{document}

\title{Time-dependent thermal transport theory} 
\author{Robert Biele} 
\email{r.biele02@gmail.com}
\affiliation{Nano-Bio Spectroscopy Group and European Theoretical Spectroscopy Facility (ETSF), Universidad del Pa\'is Vasco, E-20018 San Sebasti\'an, Spain}

\author{Roberto D'Agosta} 
\email{roberto.dagosta@ehu.es}
\affiliation{Nano-Bio Spectroscopy Group and European Theoretical Spectroscopy Facility (ETSF), Universidad del Pa\'is Vasco, E-20018 San Sebasti\'an, Spain}
\affiliation{IKERBASQUE, Basque Foundation for Science, E-48013, Bilbao, Spain}
 
\author{Angel Rubio}
\email{angel.rubio@ehu.es}
\affiliation{Nano-Bio Spectroscopy Group and European Theoretical Spectroscopy Facility (ETSF), Universidad del Pa\'is Vasco, E-20018 San Sebasti\'an, Spain}
\affiliation{Max Planck Institute for the Structure and Dynamics of Matter, Hamburg, Germany}

\date{\today}
\begin{abstract}
Understanding thermal transport in nanoscale systems presents
important challenges to both theory and experiment. In particular, the
concept of local temperature at the nanoscale appears difficult to
justify. Here, we propose a theoretical approach where we replace the
temperature gradient with controllable external black-body
radiations. The theory recovers known physical results, for example
the linear relation between the thermal current and the temperatures
difference of two black-bodies. Furthermore, our theory is not limited
to the linear regime and goes beyond accounting for non linear effects
and transient phenomena. Since the present theory is general and can
be adapted to describe both electron and phonon dynamics, it provides
a first step towards a unified formalism for investigating thermal and
electronic transport.
\end{abstract}
\maketitle

The wide research field of energy transport in nanoscale systems is very
active: groundbreaking advances in understanding the physics of this important
phenomenon have been made in recent years \cite{Volz2009,Dubi2009,Li2012},
e.g., the measurement of the quantum of thermal conductance
\cite{Rego1998,Schwab2000}, thermal quantum rectifiers \cite{Zhu2014}, breaking
of the Fourier's law \cite{Dubi2009}. Over time, different theoretical
approaches have been proposed ranging from the Landauer's theory of electrical
and thermal transport
\cite{DiVentra2008,Volz2009,Nolas2001,Chen2005b,Guedon2012}, molecular dynamics
\cite{Lee2007a}, quantum and classical Boltzmann equation \cite{Cahill2003},
non-equilibrium Green's function formalisms
\cite{Galperin2007,Wang2008,Saha2011,Stefanucci2013,Sanchez2013,Lopez2013}, to the theory of open
quantum systems \cite{Gardiner2000,Breuer2002,Biele2012,Biele2014} to point out
just a few. This activity is justified by the range of possible applications
from thermoelectric energy conversion, heat dissipation, kinetics of chemical
reactions, to a very new and more fundamental definition of thermodynamical
equilibrium \cite{Nolas2001,Ponomarev2011,Volz2009}, to name just a few. It is
a common everyday experience that two macroscopic bodies in contact with each
other equilibrate in the long-time limit to the same temperature.
Microscopically, equilibration means that there is not a net energy flow
between the two bodies but energy exchanges, in form of small fluctuations, are
still present. Since the direction of the energy flow is determined by the sign
of the difference between the temperatures of the bodies, we conclude that the
absence of an energy flow implies the two bodies have the same temperature.
This law of thermodynamics provides an operative definition of the temperature
difference. What makes thermal transport at the nanoscale a difficult
theoretical problem is that the very basic fundamentals of standard
thermodynamics cannot be applied, and the idea of thermalization needs
revisiting. In the past, attempts have been made to introduce a position
dependent temperature \cite{Zubarev1994}, but they do not provide a
satisfactory definition of local temperature. Indeed, the concepts of local
Hamiltonian, useful to define a local energy density, local thermal current,
and local nano-scale thermal gradient are not uniquely defined.
Recently, a way out, restricted to small thermal gradients, has been put
forward using an effective gravitational field, as originally proposed by
Luttinger \cite{Luttinger1964}, that mimics the effect of the temperature
gradient \cite{Eich2014,Eich2014a}. Here, we propose an alternative approach where the
temperature field is established by two or more blackbodies of known thermal
properties. Besides thermal transport at the nano-scale, this theory can be
used to understand energy transport in cold atoms, biological or optical
systems. We first lay down the basic formalism and then consider a simple
one-electron model system. However, our theory can be combined with the
general framework of time-dependent current-density functional theory (TDCDFT)
\cite{Marques2006,Ullrich2012,Giulianivignale} to consider also many-body
systems. More important, our approach is not restricted to the linear response
or weak coupling regimes, and we can easily investigate the interesting cases
of both strong coupling - recovering the Kramers' turnover \cite{Hanggi1990} -
and large temperature gradients. Finally, we have access to the full dynamics
of the system, therefore we can investigate transient regimes, usually
unaccessible to other formalisms. Last but not least, our theory can be applied
to investigating the phonon thermal transport. In this respect, it could be
seen as a first step towards a unified ab-initio formalism for thermal and
electric transport.

A blackbody, according to its original definition \cite{Kirchhoff1860}, is a
macroscopic object that absorbs all the radiation impinging on it, and in
thermal equilibrium it emits electromagnetic radiation according to the Planck's
law, whose spectrum is determined solely by the temperature of the blackbody
and not by its shape or composition \cite{Planck1901,Planck1900}. If we connect
a blackbody with a cavity made of perfect reflecting walls, the radiation
inside the cavity thermally equilibrates with the blackbody radiation, and any
object in this cavity will also thermalize. By changing the temperature of the
blackbody we control both the temperature inside the cavity and that of the
body. When thermal equilibrium is reached, at any point in the cavity the
electromagnetic radiation follows Planck's law with the temperature of the
blackbody. Ultimately we can extract the local temperature from the observation
of the energy radiation and this serves us in the following
to construct a formalism for thermal transport. Our thermometer, or thermal
source, is indeed a blackbody which radiates according to its temperature.

To begin with, we consider an electronic system coupled to any
strength to the black-body radiation and weakly to the free field of
an external environment. The dynamics of the black-body radiation is
determined solely by the blackbody itself. Here, we assume the
macroscopic parameters of the blackbody to be constant in time and
treated classically. At the same time, to allow for relaxation and 
to mimic the experimental set-ups, the system is embedded in a
bosonic environment, kept at constant temperature $T_E$, with which
the system can exchange energy via spontaneous and stimulated emission
or absorption. The total Hamiltonian where we treat the
environment quantum-mechanically (we set
$\hbar=e=m=1/(4\pi\epsilon_0)=1$), is
\begin{equation}
\hat H_T = \sum_{i=1}^N \left[\frac{\left( \hat{\bold p}_i -  
\hat{ \bold A}_F (\hat{\bold r}_i) -  { \bold A}_{\mathrm{BB}}( \hat{\bold r}_i,t)
 \right)^2}{2} +U(\hat{\bold r}_i,t)\right]+\hat H_F,
\end{equation}
where
$\hat H_F=\sum_{\bold k,s} \omega_{\bold k}\hat b^\dagger_{\bold
  k,s}\hat b_{\bold k,s} $
is the Hamiltonian of the free field of the environment and
$\hat{ \bold A}_F$ its corresponding vector potential. In addition,
$U$ is an external potential and ${ \bold A}_{\mathrm{BB}}$ describes
the electromagnetic radiation emitted by the blackbody sources. We
neglect the direct interaction of the free field with the black-body
radiation. This can be justified by observing that the photon-photon
interaction is small, consequently the contributions of the terms
where, e.g., a photon from the blackbody scatters with the free field
and then is absorbed by the system, are negligible.  Therefore, we
write $\hat H_T = \hat H_{S}(t) +\hat I_{S,F} +\hat H_F$ where we have
included the system-blackbody interaction in the system Hamiltonian,
$\hat H_S = \sum_{i=1}^N \left(\hat{\bold p}^2_i/2 +U(\bold r_i,t) -
  {\bold A}_{\mathrm{BB}} (\bold r_i,t)\cdot \hat{\bold p}_i\right)$,
and the quadratic term ${\bold A}_{\mathrm{BB}}^2$ in the external
potential $U$. The free vector-field is written as
$\hat{\bold{A}}_F(\bold r) = \sum_{\bold k,s} p_{\bold k,s} \left(
  \hat b_{\bold k,s} \exp(i\bold{k}\bold r) + \hat b^{\dagger}_{\bold
    k,s} \exp(-i\bold{k}\bold r) \right)
\bold{\epsilon}_{\bold{k},s}$,
where $p_k=\sqrt{2\pi/(\omega_k V)}$ and $\hat b_{\bold k}^{\dagger}$
is the creation operator for a free-field mode with polarization
direction $\bold{ \epsilon}_{\bold{k},s} $. Finally, by exploiting the
field expansion
$\hat\Psi = \sum_{\alpha} \hat c_{\alpha} (t) \phi_{\alpha}(r)$, the
system-environment interaction can be written, in the Coulomb gauge as
$\hat I_{S,F} = \sum_{\alpha,\beta,\bold{k},s}
g_{\alpha\beta,\bold{k},s}\:\hat c_{\alpha}^{\dagger} \hat c_{\beta}
\left(\hat b_{\bold k, s} + \hat b_{\bold k, s}^{\dagger} \right) $,
where we have defined,
$g_{\alpha\beta,\bold{k},s} = i p_{\bold k}\int \mathrm{d} V
\phi_{\alpha}^{*}(\bold{r}) \bold{ \epsilon}_{\bold k ,s} \cdot
\bold{\nabla} \phi_\beta ( \bold r) $.
Here, we have assumed that the system size is small in comparison to
the wavelengths of the electromagnetic field (dipole approximation)
\footnote{Going beyond the dipole approximation is straightforward,
  but for the sake of simplicity we will use it in this
  work. Furthermore, we neglected the quadratic terms in the
  interaction of the system with the free-field as we assume that the
  coupling to the environment is small.}.  Furthermore, the operators
$\hat c^\dagger_\alpha$ create the $\alpha$'s energy eigenstate of the
initial Hamiltonian $H_S(0)$. These energy eigenstates will serve as a
natural basis set for our following considerations. Naturally, the (local) 
density of states associated with the eigenstates of this
Hamiltonian defines how efficient the coupling between the system and the 
black-body radiations is \cite{Stefanucci2013,DiVentra2008}.

As we are interested only in the system dynamics, we will examine the dynamics of the expectation value of the one-particle density operator
$f_{\alpha\beta} = \langle\hat c^{\dagger}_{\alpha} \hat c_\beta\rangle$,  easily derived from the equation of motion for
$\hat c^{\dagger}_{\alpha} \hat c_\beta$, $i\partial_t f_{\alpha\beta} =
\langle[\hat H_S(t),\hat c^{\dagger}_{\alpha} \hat c_\beta]\rangle 
	+\sum_{\gamma,\bold k} [g_{\beta,\gamma,\bold k} 
	( \langle \hat b_{\bold k} \hat c^\dagger_\alpha \hat c_\gamma\rangle +\langle \hat b^\dagger_{\bold k} \hat c^\dagger_\alpha \hat c_\gamma\rangle) 
- g_{\gamma,\alpha,\bold k}( \langle \hat b_{\bold k} \hat c^\dagger_\gamma \hat c_\beta\rangle + \langle \hat b^\dagger_{\bold k} \hat c^\dagger_\gamma \hat c_\beta\rangle)]
$,
where we included the spin index $s$ in the components
of $\bold k$. 
Its solution requires the knowledge of the dynamics of
$\hat b_{\bold k} \hat c^\dagger_\alpha \hat c_\beta$,
\begin{eqnarray}
	\label{eq:bcc} 
		i \partial_t \hat b_{\bold k}\hat c^{\dagger}_{\alpha} \hat c_\beta&=& \left[\hat H_S,\hat b_{\bold k}\hat c^{\dagger}_{\alpha} \hat c_\beta\right]
+\omega_{\bold k}\hat b_{\bold k}\hat c^{\dagger}_{\alpha} \hat c_\beta\nonumber\\
&&+ \sum_{\gamma,\delta,\bold l} g_{\gamma\delta,\bold l}\left[ \hat c^\dagger_\alpha \hat c_\beta \hat c^\dagger_\gamma \hat c_\delta \delta_{\bold k,\bold l} \right.\\
&&+\left. \left\{ \hat b_{\bold k}(\hat b_{\bold l} 
+\hat b^\dagger_{\bold l})-\delta_{\bold k,\bold l}\right\}\left(\hat c^\dagger_\alpha \hat c_\delta \delta_{\gamma\beta}
- \hat c^\dagger_\gamma \hat c_\beta \delta_{\delta\alpha} \right) \right]\nonumber.
\end{eqnarray}
A similar equation holds for $\hat b^\dagger_{\bold k} \hat c^\dagger_\alpha \hat c_\beta$.
These equations of motion are not in a closed form,
and any attempt to solve them by investigating the dynamics of the operators
appearing on the right hand side leads to an infinite hierarchy of equations.
For this reason, we decouple the dynamics of the system and the field, i.e., we
assume that $ \langle\hat b_{\bold k} \hat b^\dagger_{\bold l} \hat
c^\dagger_\alpha \hat c_\beta\rangle \approx \langle\hat b_{\bold k} \hat
b^\dagger_{\bold l}\rangle\langle \hat c^\dagger_\alpha \hat c_\beta\rangle.$
With this approximation Eq. (\ref{eq:bcc}) is
solved with a standard integration technique. Furthermore, by using that the
initial-state correlation vanishes, $\langle\hat b^\dagger_{\bold k}\hat
c^{\dagger}_{\alpha} \hat c_\beta\rangle(0)=\langle\hat b_{\bold k}\hat
c^{\dagger}_{\alpha} \hat c_\beta\rangle(0)=0$ and that the environment is in
thermal equilibrium,
$n(\omega_{\bold k},T_E)= \sum_{\bold l}\langle\hat
b_{\bold k}(\hat b_{\bold l} +\hat b^\dagger_{\bold l})-\delta_{\bold k,\bold
l}\rangle$, where $n(\omega,T)$ is the Planck's distribution at energy $\omega$
and temperature $T$, we arrive at a non-Markovian master equation
\begin{equation}
	\begin{split}
\label{eq:master_eq} 
i\partial_t f_{\alpha\beta} =& \left\langle\left[ \hat f,\hat H_S\right]\right\rangle_{\alpha\beta}\\
&+ i\sum_{\bold k,\bold l}\int_{t_0}^t\mathrm{d}\tau\:C_{\bold k,\bold l}(\tau,t)\left\langle\left[ \hat U(\tau,t) \hat f(\tau) \hat V_{\bold l} \hat U^\dagger(\tau,t)\hat V_{\bold k}\right.\right.\\
&-\hat U(\tau,t)\hat V_{\bold l} \hat f(\tau) \hat U^\dagger(\tau,t)\hat V_{\bold k}\\
&- \hat V_{\bold k} \hat U(\tau,t) \hat f(\tau) \hat V_{\bold l} \hat U^\dagger(\tau,t)\\
&\left.\left. +\hat V_{\bold k} \hat U(\tau,t) \hat V_{\bold l} \hat f(\tau) \hat U^\dagger(\tau,t) \right]\right\rangle_{\alpha\beta} \\
&-2\int_{t_0}^t\mathrm{d}\tau\sum_{\substack{\gamma,\delta,\sigma,\zeta
\bold k,\bold l}} V^*_{\delta\gamma,\bold k}V^*_{\sigma \zeta,\bold l}\sin(\omega_{\bold k}(t-\tau))\\
&\times\{ \delta_{\beta\gamma}U_{\delta\alpha}(\tau,t) \langle\hat c^\dagger_\alpha \hat c_\delta \hat c^\dagger_\sigma \hat c_\zeta\rangle(\tau)\\
& -\delta_{\delta\alpha}U_{\beta\gamma}(\tau,t) \langle\hat c^\dagger_\gamma \hat c_\beta \hat c^\dagger_\sigma \hat c_\zeta\rangle(\tau)\}.
\end{split}
\end{equation}
Here, we have introduced the bath correlation function $C_{\bold k,\bold l}(\tau,t) = \langle\{\hat b_{\bold k}(\tau) +\hat b^\dagger_{\bold k}(\tau)\}\{\hat b_{\bold l}(t) +\hat b^\dagger_{\bold l}(t)\}\rangle
	= (n(\omega_{\bold k},T)+1)e^{-i\omega_{\bold k}(\tau-t)}+n(\omega_{\bold k},T)e^{i\omega_{\bold k}(\tau-t)}$,
and defined $V^*_{\beta\alpha,\bold k}=-ig_{\alpha
\beta,\bold k}$ and $\hat U(\tau,t)=\hat
T_{+}e^{-i\int_{\tau}^t\mathrm{d}t^\prime \hat H_S(t^\prime)}$, where $\hat
T_{+}$ is the time-ordering operator. The former equation describes a system
under the influence of a classical black-body radiation, where the system can
dissipate to or gain energy from the environment.
It is known that for the system described by this master equation the
detailed balance of the power spectrum of the correlation function
$C_{k,l}(\omega)=\int_{-\infty}^{\infty}\exp(-i\omega
(t-\tau))C_{k,l}(t,\tau)dt$, is the minimum requirement to reach thermal
equilibrium with the free field. One can easily check that the detailed balance
condition is satisfied by our correlation function. Then,
without any external sources, the system evolving according to
Eq. (\ref{eq:master_eq}) reaches thermal equilibrium with the free field
radiation. On the other hand, when external sources or driving potentials are
present, the system does not reach any thermal equilibrium in 
general. However, we expect the system to reach a steady state regime
in the long time limit. This expectation is rooted in the observation that
stimulated and spontaneous emissions grow when the system is strongly driven
until a balance between the energy absorbed from the external fields and that
emitted is reached.

\begin{figure} 
	\includegraphics[width=8.cm]{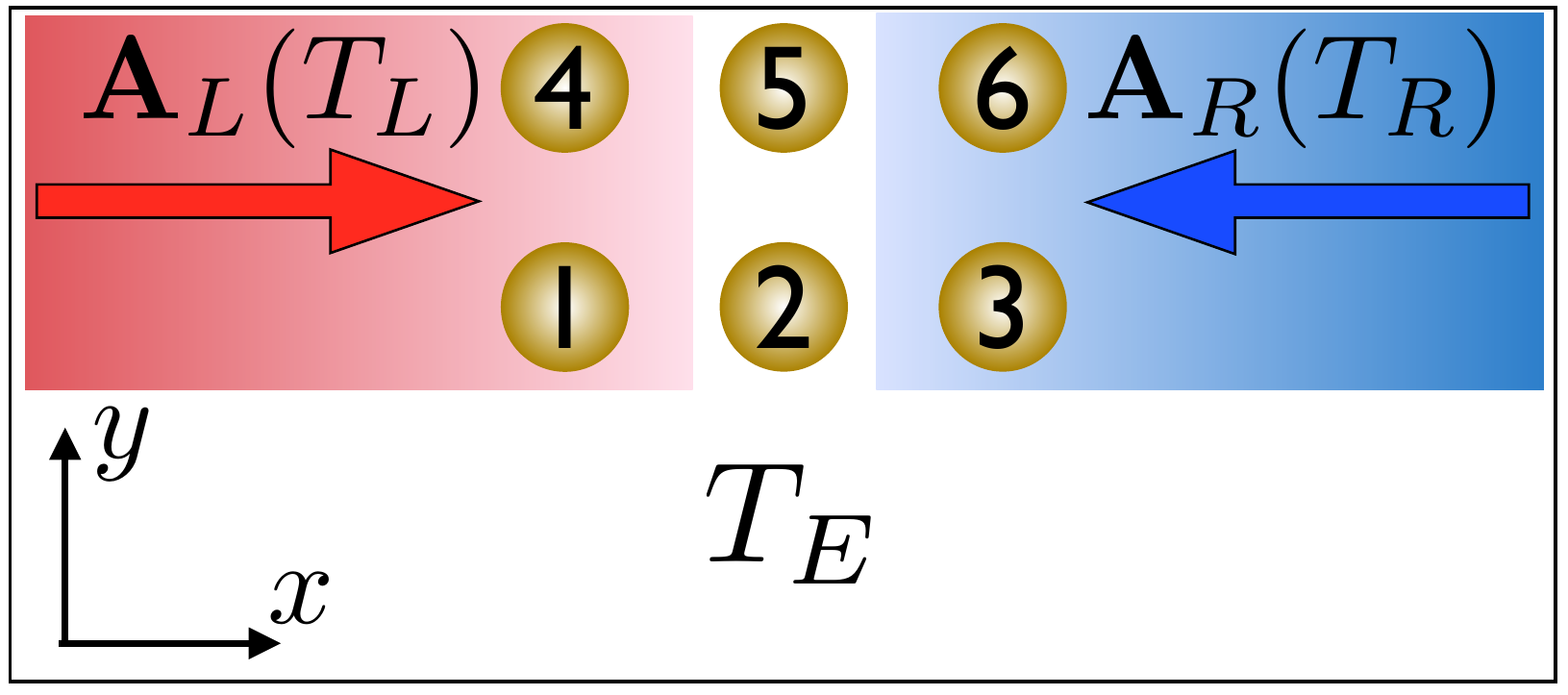} 
	\caption{(Color online) Sketch of the set-up under consideration. The tight
binding sites are labeled by 1 to 6 and are connected on the left and right
side to two blackbody radiations at different temperature, $T_L$
and $T_R$. This system is embedded in an environment at temperature $T_E$.}
	\label{tight-binding} 
\end{figure}

In the following we will demonstrate and test the theory on a model
system of fundamental importance, namely we will study heat transport
induced by the black-body fields in a small two dimensional spin-less
tight-binding system sketched in Fig. \ref{tight-binding}. First of
all we check whether known results are reproduced. For this, we prove
that the system relaxes in the long-time limit to its thermal
equilibrium, and find that our system also shows a `Kramers
turnover'-like behavior \cite{Hanggi1990, Velizhanin2008} as
expected. The tight-binding sites are labeled by the numbers one to
six, and they are connected via nearest-neighbor hopping. Here,
$\bold A_L$ and $\bold A_R$ represent the electromagnetic field from
two blackbodies at positions $x=-\infty$ and $x=+\infty$ with
temperatures $T_L$ and $T_R$, respectively. In addition, the system is
embedded in an environment at temperature $T_E$.

As these blackbodies are far away from the system, their fields are a
superposition of plane waves traveling in positive (or negative)
$x$-direction, weighted according to their temperature,
$\bold A_{L,R}(r,t) = {\bf E_0} \int \mathrm {d} \Omega \sqrt{\Omega
  n(\Omega,T_{L,R})}\sin\left(\Omega t \pm kx +\phi (\Omega)\right), $
where ${\bf E_0}=E_0 {\bf e}_y$, and $E_0$ is the strength of the
corresponding electric field, ${\bf e}_y$ is the unit vector in the
$y$ direction, and $\phi(\Omega)\in [0,2\pi]$ are uncorrelated random
phase factors. Here, one might use more realistic, and also
complicated, models for the correlation of thermal radiation
\cite{Donges1999,Bertilone1996}. This thermal radiation enters through
the Peierls transformation of the hopping parameter
$T^A_{ij} =T_{ij} \: \exp\left(-i
  \int_{\bold{R}_i}^{\bold{R}_j}\mathrm{d}\bold r \cdot (\bold
  A_L(r,t)+\bold A_R(r,t))\right)$,
into the tight-binding hamiltonian,
\begin{equation}
	\hat H_S=\sum_{\langle i,j\rangle} T^A_{ij} \hat c_i^\dagger \hat c_j. 
	\label{hamiltonian_tb}
\end{equation}
Here, the operator $\hat c^{\dagger}_i$ creates an electron at site
$i$ with position $\bold R_i$, and we assume a single electron to be
present in the whole system. The sum $\langle i,j\rangle$ denotes
summation over nearest neighbors only.  The vector potentials
$\bold A_L$ and $\bold A_R$ couple to the most leftwards and
rightwards sites, respectively, and will introduce a local temperature
gradient in the system. Note that in general the potentials penetrate
into the system, however for this model system, we assume the external
radiation is rapidly screened. For example, the core
  electrons in the tight-binding sites can be responsible for this
  screening.

\begin{figure}[ht!] 
	\includegraphics[width=8.cm]{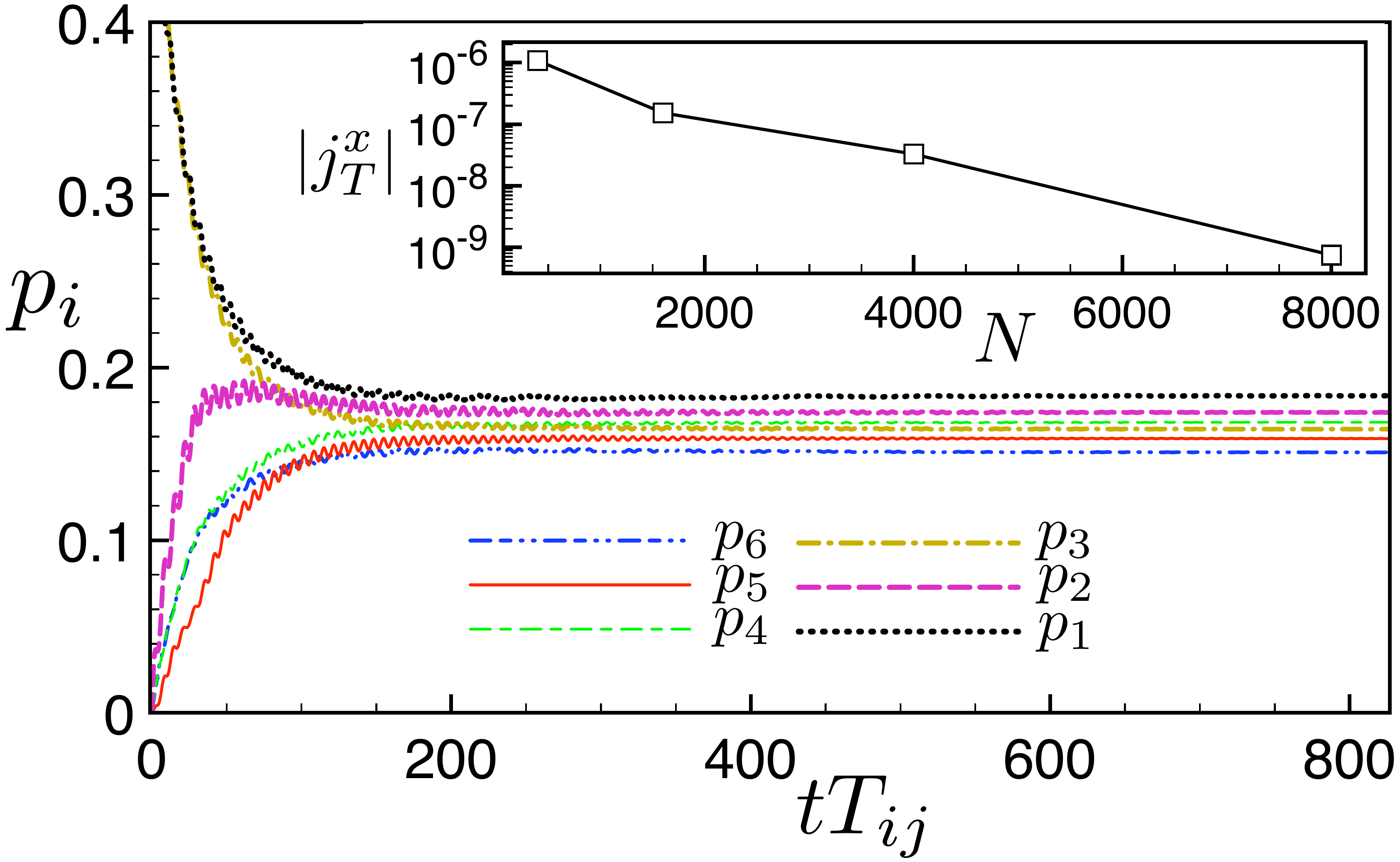} 
	\caption{(Color online) Relaxation dynamics of the occupation probabilities of the eigenstates of the unperturbed Hamiltonian (\ref{hamiltonian_tb}). This results have been obtained by averaging over $4\,000$ realizations of the stochastic noise. The inset shows the vanishing energy current through the system when the number of runs of the stochastic process, $N$, increases.
} \label{relaxation} 
\end{figure}

The coupling to the environmental degrees of freedom is described by a master
equation. For the coupling of an electronic system to the electro-magnetic
field of the environment, the coupling operator $\hat V$ and bath-correlation
function $C(\tau,t)$ can be derived from first principles by assuming the
system to be embedded in a cavity at temperature $T_E$ \cite{Biele2014}. The
power spectrum, essentially the Fourier transform of the bath-correlation
function, is given by $ C_{\mathrm{env}}(\omega)={4|\omega|^3}/{c^3}
\left[n(|\omega|,T_E)+\theta(-\omega)\right]$ for $|\omega|<\omega_{c}$ where
$\theta(\omega)$ is the Heaviside step function and $\omega_c$ is a cutoff
frequency determined by the dimensions of the system. For $|\omega|>\omega_c$,
the power spectrum is set to vanish. This cutoff is necessitated by the
assumption made in the dipole approximation that the electromagnetic field is
uniform in the region of space occupied by the system. The corresponding
coupling operator, entering the master equation, is given by $\hat V = -
\sum_{i,j} \bold{u} \cdot \langle W_i | \bold{r} | W_j \rangle\, \hat
c_i^{\dagger} \hat c_j,$ where $|W_i\rangle$ is the single-particle state
localized at site $i$. As we are interested in the steady-state energy current
through the system, we simplify the memory kernel of the non-Markovian master
equation (\ref{eq:master_eq}) by setting $\int_0^t \mathrm{d}\tau \hat
f(\tau)[...]\rightarrow \int_{-\infty}^\infty \mathrm{d}\tau \hat f(t)[...]$.
This approximation does not change the long-time limit of the equation of
motion and hence is suitable to study steady-state dynamics \cite{Biele2012}.
In order to calculate the energy transport one has to define, via the
continuity equation, an energy current in the system. With the local
energy-density operator, $ \hat h_i = \frac{1}{2}\sum_{\langle j\rangle}\big(
T^A_{ij}\hat c^\dagger_i \hat c_j +\mathrm{h.c.} \big)$, one can define the
total current in the $x$-direction via $ \hat j^x_T = - \dot {\hat h}_1 - \dot
{\hat h}_4 + \dot {\hat h}_3 + \dot {\hat h}_6$, where $\dot{\hat O}=i[\hat
H_S,\hat O]+\partial_t \hat O$. As a first test of the novel thermal transport
theory we examine the relaxation dynamics of the tight-binding system driven by
a left and right blackbody radiation kept at temperature $k_B T=10\,a.u.$, the
same temperature as the environment. We choose $T_{ij} = 0.5\,a.u.$ as an
energy-scale for the system, and for the coupling to the environment we set
$\gamma = \sqrt{2/(\pi c^3)}=0.05$. For the energy scale of the black-body
radiations, we have normalized both left and right radiation with the
time-averaged Poynting vector, $\langle S\rangle$, set to $15/(\mu_0\cdot c)$.
In Fig. \ref{relaxation} the dynamics of the occupation probabilities of the
eigenstates of the Hamiltonian (\ref{hamiltonian_tb}) in the one-electron
sector is shown for the unperturbed system (zero black-body field). One can see
that the system relaxes towards a steady state. The steady state is
characterized by zero energy transport in the system, which can be seen in the
inset of Fig. \ref{relaxation}.

\begin{figure}
	[ht!] 
	\includegraphics[width=8.cm]{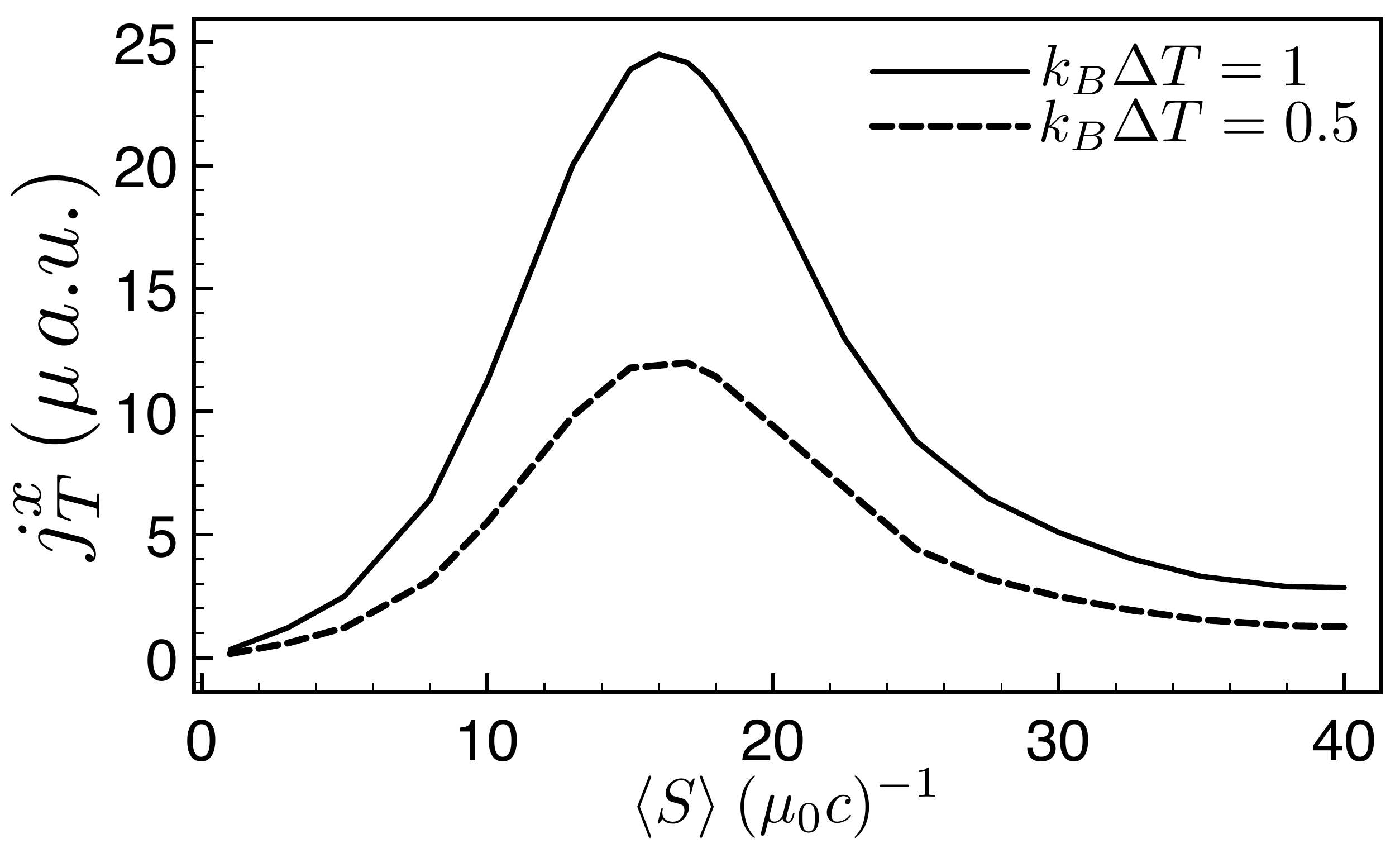} 
	\caption{Dependence of the steady-state energy current on the coupling strength $\langle S\rangle$ for $k_BT_E=10$ and $N=4\,000$. A turnover in the energy current can be observed.
} \label{turnover_fig} 
\end{figure}

\begin{figure}[ht!] 
	\includegraphics[width=8.cm]{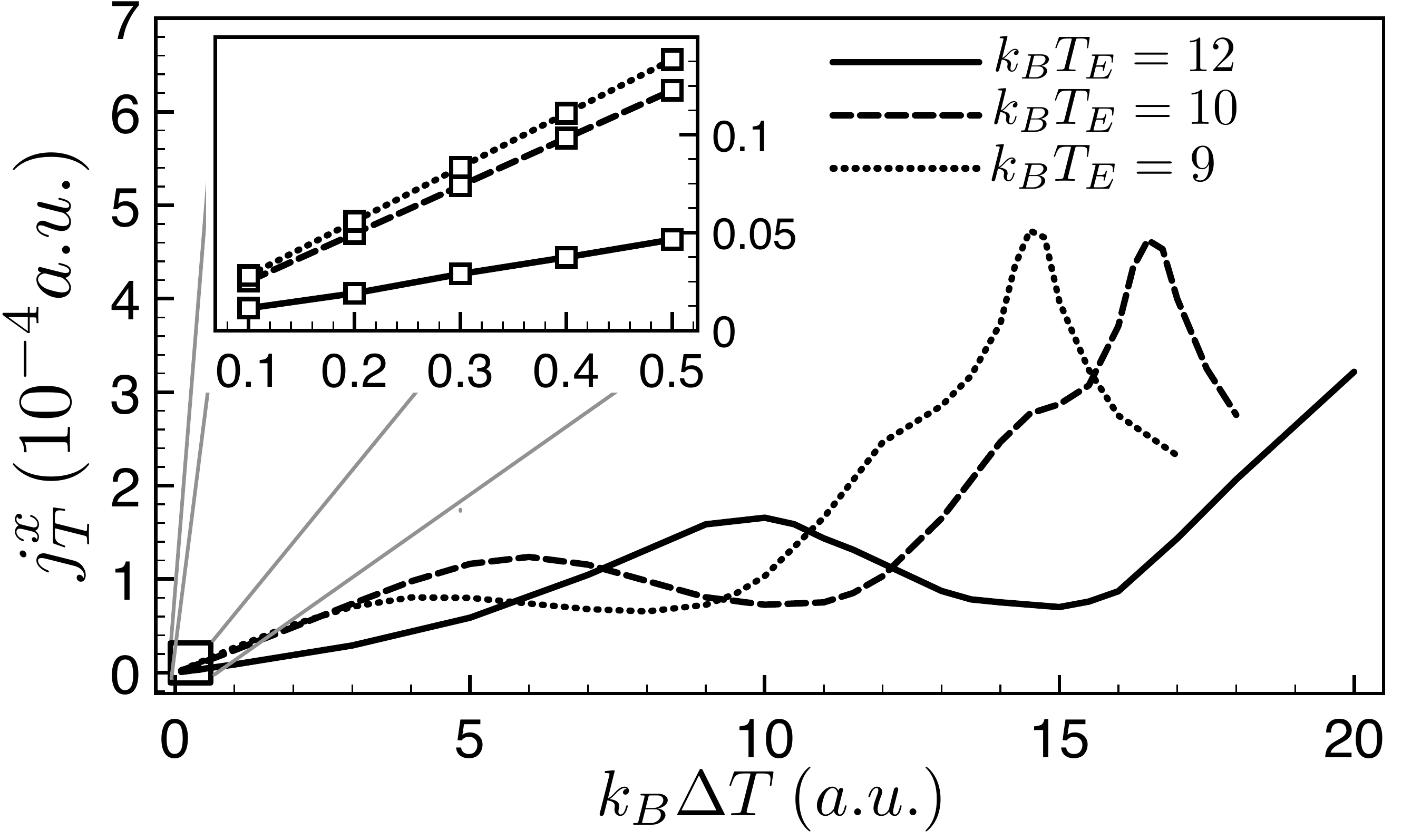} 
	\caption{Energy current in the tight-binding system of Fig.
\ref{tight-binding} vs. temperature gradient introduced by the black-body
radiations around the environmental temperature $T_E$. 
For this plot, we have used $\langle
S\rangle=15/\mu_0 c$ close to the maximum current of Fig.~\ref{turnover_fig},
and $N=4\,000$.}
\label{energy-current} 
\end{figure}

To further verify the theory we show one finds the expected turnover behavior as seen in Fig.~\ref{turnover_fig} \cite{Hanggi1990, Velizhanin2008}. The turnover can be understood by considering that for small energy flux, when increasing the flux, more states are excited and can contribute to transport. On the other hand, at large fluxes, some of the states are fully occupied and are not able to contribute to transport anymore. At intermediate energy fluxes, a peak of the energy current must be achieved. In general, this behavior cannot be described by perturbative theories for thermal transport such as the Redfield theory \cite{Velizhanin2008}. We have set the temperature gradient to $\Delta T=(T_L-T_R)/2$. 
In Fig.~\ref{energy-current} the energy current is plotted versus the
introduced thermal gradient, $\Delta T$, from the black bodies. A linear
dependence on the thermal gradient can be found for $\Delta T \ll T$ (see
inset). At large $\Delta T$, a maximum in the energy current is reached. This
implies that the system can sustain up to a maximum energy flow fixed the
external conditions. Also, the position of the maximum of the energy current
shifts to the right by increasing the temperature of the environment. This might be
relevant for technological applications since the maximum efficiency can tuned
according to the working conditions.

In conclusion, we have presented a novel theoretical framework to investigate thermal and energy transport, where the thermal imbalance
in the system is introduced by two classical black body radiations. Our theory also includes a dissipative environment, where the system
can gain energy from or dissipate to, in order to mimic the quantum nature of the photons. 
Due to the latter, we include the fundamental concept of thermal relaxation of the system, which
is not included in other thermal transport theories. The theory can also be used in different set-ups, e.g., we can consider one
blackbody only, to adapt to different experiments. Finally, as our formalism relies on the knowledge of the external vector potential,
we can foresee that its combination with the powerful techniques of TDCDFT will provide a novel ab-initio tool to study thermal
transport in many-body systems, and possibly pave the way to define a local temperature.
In addition, by utilizing a recent development in combining Time-Dependent Density Functional theory and quantum electro-dynamics \cite{Ruggenthaler2014}, we can go beyond the mean-field description of the environment.
 Moreover, within the same formalism we can investigate phonon thermal transport, thereby combined with
the TDCDFT, our model provides a unified way to investigate ab-initio electrical and thermal transport beyond linear response.
Non-linear regimes are important since in seeking for, e.g., the maximum efficiency of a thermoelectric energy converter, we might need
to go beyond the standard linear response \cite{DAgosta2013}. Since our formalism is fully dynamic we have direct access to transient
regimes, to how the steady state is approached, and to whether or not this steady state is unique or depends on the history of the
system \cite{Stefanucci2007}.

\acknowledgments 
We acknowledge financial support by the European
Research Council Advanced Grant DYNamo (ERC-2010-AdG-267374), FUN-EMAT
(FIS2013-46159-C3-1-P), Grupo Consolidado UPV/EHU del Gobierno Vasco
(IT578-13), and NANOTherm (CSD2010-00044). R. B. acknowledges the
financial support of Ministerio de Educacion, Cultura y Deporte
(FPU12/01576), useful discussions with C. Timm, and the hospitality of
his group at the Technische Universit\"at Dresden.

\bibliography{library.bib}

\end{document}